\documentclass{ws-mplb}
\usepackage{amsfonts,amssymb,amsmath}
\usepackage{graphicx}
\usepackage[super,sort,compress]{cite}
\begin{document}
\markboth{Matej Hudak, Ondrej Hudak}{On composites of the type: metal - dielectrics and superconductor - dielectrics.}
\catchline{}{}{}{}{}
\title{ \Large On composites of the type: metal - dielectrics and superconductor - dielectrics.}
\author{Matej Hudak}

\address{Stierova 23\\
 SK-040 23 Kosice\\
hudakm@mail.pvt.sk}

\author{Ondrej Hudak}

\address{Department of Aviation Technical Studies, Faculty of Aerodynamics, \\ Technical University Kosice,  Rampova 7, SK - 040 01 Kosice, Slovak Republic\footnote{Corresponding author}\\
hudako@mail.pvt.sk}

\maketitle

\begin{history}
\received{(12-st December 2017)}
\revised{(Day Month Year)}
\end{history}

\begin{abstract}
Composites of the type: metal - dielectrics and superconductor - dielectrics are studied in the quasistatic approximation.  The dielectric response is described by the spectral function $G(n,x)$, which contains effects of the concentration x (of metallic resp. superconductive particles) on the dielectric function,and effects of the shape. The parameter n plays the role of the depolarisation factor for dielectric materials, in metals it is a factor which includes effects like shape, and a topology of the composite. There exists a percolation transition at $ x_{c}= \frac{1}{3} $ which leads to a metallic-like for the composite with the concentration $ x > x_{c}$. At low frequencies divergence with frequency remains even when there are present dielectric particles above the percolation concentration. In superconductor case the spectral function $G(n,x)$ may include also Josephson junction effects. We  assume in both cases of composites two types of spheroidal particles, metal (superconducting) ones and dielectric ones.  A dielectric function is constant in both cases for the dielectric material, and a dielectric function for the metal and for the superconductor are used with well known form  for metals and a classical superconductor. A percolation transition at $ x_{c}$ leads to a metallic-like absorption for the composite with $x>x_{c}$. Note that at low frequencies divergence in frequency remains even when there are present dielectric particles above $x_{c}$. Below the percolation threshold dielectric properties are modified by metalic particles. We obtain at very low temperatures and low concentrations x of the superconductor the effective dielectric constant. The absorption part is zero in our simple case. The real part of the dielectric function increases with the concentration of the superconducting spheres. The frequency dependence is quadratic, it gives low frequency tail.
\end{abstract}

\section{On composites of the type: metal - dielectrics.}

Qusistatic approximation in description of the dielectric response of the composite of the type metal - dielectrics is assumed to be a valid approximation\cite{RO1}$^{-}$\cite{H} . The dielectric response can be described by the spectral function $G(n,x)$ which contains effects of the concentration on the dielectric complex function and effects of the shape. The parameter n plays the role of the depolarisation factor for dielectric materials, in metals it is a factor which includes effects like shape, and a topology of the composite. We  assume here two types of spheroidal particles, metal ones and dielectric ones. A dielectric function is assumed to be constant for the dielectric material, and a dielectric function for the metal is used in well known form  for metals.

The effective dielectric function of the composite is assumed to have the form:
\begin{equation}
\label{1}
\epsilon_{eff} = \epsilon_{D} (1 - x \int_{0}^{1}\frac{G(n,x)}{t-n} dn ),
\end{equation}
where:
\begin{equation}
\label{2}
t \equiv \frac{\epsilon_{D}}{\epsilon_{D}-\epsilon_{M}}, 
\end{equation}
and where $ \epsilon_{D} $ is a dielectric function (constant in this case) for the dielectric material, and where $\epsilon_{M} $ is a dielectric function for the metal, with well known form:
\begin{equation}
\label{3}
\epsilon_{M} = 1 + \chi_{L} + ( \frac{i\frac{\sigma_{0}}{\epsilon_{0}\omega }}{1 - i \omega  \tau } ).
\end{equation}
Here $ \tau $ is a relaxation time for a given metal electrons, $ \omega $ is a frequency, $ \epsilon_{0} $ is a static dielectric constant in metal, the constant $ \sigma_{0} = \frac{e^{2} n_{e} \tau }{m}  $, where $e$ is a charge of the electron, $ n_{e} $ is a concentration of conductivity electrons and m is its mass. 

In the low frequency limit, $ \omega \tau << 1 $, using the form of the spectral function for the MGT limit\cite{RO1}$^{-}$\cite{H} we find that the effective dielectric constant absorption part $\epsilon_{eff}^{,,}$ of the composite is given by:
\begin{equation}
\label{4}
\epsilon_{eff}^{,,} = \frac{\sigma_{0}}{\epsilon_{0}\omega } (1 - \frac{3(1-x)}{2}) = \frac{\sigma_{0}}{\epsilon_{0}\omega }. \frac{(x-x_{c})}{2}  .
\end{equation}
Thus we conclude, starting from this simple case, that there exists a percolation transition at $ x_{c}= \frac{1}{3} $ which leads to a metal absorption for
a pure metallic  composite (x=1), and to metallic-like absorption for the composite with the concentration x above the percolation threshold $x_{c}$. Note that at low frequencies
divergence in the frequency in (\ref{4}) remains present even when there are present dielectric particles (above the percolation concentration $x_{c}$ of metalic particles. Dielectric particles are not connected into an infinite cluster for $x > x_{c}$ .

\section{On composites of the type: superconductor - dielectrics.}

Qusistatic approximation in the description of the dielectric response of the composite of the type: superconductor - dielectrics material  is assumed to be valid approximation here.
 The dielectric response for this composite will be described again by the corresponding spectral function $ G(n,x)$ which contains effects of the concentration on the dielectric complex function, and effects of the shape. Parameter n  in superconductors is a factor which includes effects like shape of the particles and topology of the
composite. It may include also Josephson junctions effects. We  again assume two types of spheroidal particles, superconducting ones and dielectric ones. The effective dielectric function $\epsilon_{eff}$ is assumed to have again the form:
described above:
\begin{equation}
\label{1'}
\epsilon_{eff} = \epsilon_{D} (1 - x \int_{0}^{1} \frac{G(n,x)}{t-n} dn),
\end{equation}
where:
\begin{equation}
\label{2'}
t \equiv \frac{\epsilon_{D}}{\epsilon_{D}-\epsilon_{S}} 
\end{equation}
now, and where $ \epsilon_{D} $ is a dielectric function (again constant in this case) for the dielectric material, and where $\epsilon_{S} $ is a dielectric function for the superconductor in well
known form  for the classical superconductor:
\begin{equation}
\label{3'}
\epsilon_{S} = \epsilon_{S}^{,}=1 + \chi_{L} + \frac{\pi \Delta \sigma_{n}}{\epsilon_{0}\hbar \omega^{2}  }
\end{equation}
at low temperatures well below the transition temperature from the normal phase to the superconducting phase, and for low frequencies $\omega$ well below the gap $ \Delta $ frequency. Here $ \sigma_{n} $ is the normal state constant.
The absorption part $\epsilon_{S}^{,,}$ of the dielectric constant $\epsilon_{S}$  is zero in our simple case. 

At very low temperatures and low concentrations x of the superconductor particles we obtain the effective dielectric constant $\epsilon_{eff}$ of the composite of the type: superconductor - dielectrics of the form:
\begin{equation}
\label{4'}
\epsilon_{eff} = \epsilon_{D} + x \frac{\hbar \omega^{2}  }{\epsilon_{0}\pi \Delta \sigma_{n}}.
\end{equation}
The real part of the dielectric function $\epsilon_{eff} $ increases with the concentration of the superconducting spheres. The frequency dependence is quadratic now, it gives low frequency tail which is present in many glasses and composites in general.

\section{Discussion.}

For small concentrations of the superconducting material dielectric properties are determined by the dielectric component of the composite. For large concentrations
of the superconducting component around and above the corresponding critical value of the concentration $x$ of superconducting particles, $x_{c}$, the superconducting infinite cluster (clusters) determines behaviour of the composite. Interesting region of the percolation concentration shows that topology and shape of particles are very important characteristics, there exists besides the dielectric material contribution and besides the superconducting material contribution also contribution due to topology of the composite, of the shape of
the particles. The Josephson junctions effects may be included phenomenologically into account in this region of strongly interacting superconducting particles 
in an appropriate form of the spectral function.  The effective dielectric constant is a starting point for new kind of description of experimental results in this region of concentrations using the spectral function $G(x,n)$ for a composite: metal - dielectric and for a composite: superconductor - dielectrics. Composites of this type
 are used in automotive and aviation (and other) industries,  and in the space industry,

\section*{Acknowledgement.}

Authors would like to express their sincere thanks to the late V. Dvorak from the Institue of Physics AS Czech Republic in Prague for his discussions on composites and their dielectric response.

\end{document}